\documentclass[runningheads]{llncs}
\usepackage[utf8]{inputenc}
\usepackage{graphicx}
\usepackage{amsmath}
\usepackage{multirow}
\usepackage{hyperref}
\usepackage[capitalize]{cleveref}
\usepackage{siunitx}
\usepackage[table,xcdraw]{xcolor}
\usepackage{sidecap}

\newcommand{\new}[1]{\textcolor{black}{#1}}

\begin{document}
\title{Identifying and Combating Bias in Segmentation Networks by leveraging multiple resolutions}
\titlerunning{Identifying and Combating Bias in Segmentation Networks}
\author{Leonie Henschel\inst{1} \and David Kügler\inst{1} \and Derek S Andrews\inst{2} \and Christine W Nordahl\inst{2} \and Martin Reuter\inst{1,3,4}}
\index{Henschel, Leonie}
\index{Kügler, David}
\index{Andrews, Derek S}
\index{Nordahl, Christine W}
\index{Reuter, Martin}

%
%\orcidID{0000-0002-2665-9693}}
%
\authorrunning{L. Henschel, D. Kügler, D. Andrews, C. Nordahl and M. Reuter}
%\authorrunning{Anonymous Submission}
\institute{German Center for Neurodegenerative Diseases (DZNE), Bonn, Germany \email{martin.reuter@dzne.de} \and 
%Medical Investigation of Neurodevelopmental Disorders (MIND) Institute and Department of Psychiatry and Behavioral Sciences, UC Davis, Davis, USA \and
MIND Institute and Dept. of Psychiatry and Behavioral Sciences, UC Davis, USA \and
Department of Radiology, Harvard Medical School, Boston, USA \and
A.A. Martinos Center for Biomedical Imag., Mass. General Hospital, Boston, USA}
%*********************************************************** \and
%***********************************************************}
%\institute{German Center for Neurodegenerative Diseases, Bonn, Germany  \\ \email{\{dzne\}@dzne.de} \and
%Department of Radiology, Harvard Medical School \\
\maketitle              % typeset the header of the contribution
%
%\vspace{-1.5ex}
\begin{abstract}
Exploration of bias has significant impact on the transparency and applicability of deep learning pipelines in medical settings, yet is so far woefully understudied.
In this paper, we consider two separate groups for which training data is only available at differing image resolutions. For group H, available images and labels are at the preferred high resolution while for group L only deprecated lower resolution data exist. We analyse how this resolution-bias in the data distribution propagates to systematically biased predictions for group L at higher resolutions. Our results demonstrate that single-resolution training settings result in significant loss of volumetric group differences that translate to erroneous segmentations as measured by DSC and subsequent classification failures on the low resolution group. We further explore how training data across resolutions can be used to combat this systematic bias. Specifically, we investigate the effect of image resampling, scale augmentation and resolution independence and demonstrate that biases can effectively be reduced with multi-resolution approaches.

%\keywords{Cortical Segmentation  \and Bias \and Multi Resolution.}
\end{abstract}

\section{Introduction}

Over the last years, deep learning networks have been shown to accurately segment brain MRIs to the point that they rival traditional pipelines in reliability, sensitivity and accuracy \cite{Henschel_2020,Iglesias_2021,slant2019}. However, for supervised training all networks rely on reference segmentations with the core assumption of representative training sets. In medical imaging, this assumption is often inherently violated due to limited data availability -- specifically for uncommon pathologies, age groups, treatment effects, genetic and ethnic groups as well as resolutions. There are, for example, no manual annotation or disease datasets available at leading sub-millimeter resolutions. In consequence, networks performing well in the training domain fail to generalize to distributions outside this scope \cite{Gonzalez2021}. Unfortunately, most of these effects can only be adequately addressed by acquiring more data. 

Datasets with heterogeneous resolutions may, however, be investigated by resizing available ground truth. %Downsampling of images only reduces the provided information content per voxel, resizing and comparison of images at lower resolutions is an adequate approach to generate cross-resolution training sets. 
Using this setup we can address a hitherto unexplored and important question: what effects do (systematic) resolution differences between populations or groups have on segmentation networks? Can we learn segmentation tasks across resolutions? Cross-resolution training is import, because of the trend to acquire images at submillimeter resolutions, where manual labels are not yet available.

% or are biases introduced leading to consistent errors for certain structures or downstream measures Undiscovered segmentation biases in neural networks might heavily impact future large-cohort studies specifically with regard to group separations and subsequent diagnosis or treatment decisions. %As an example, cortical gray matter thickness slowly reduces in aging and, therefore, differs significantly between adults and children. Hence, the question arises if networks trained solely on available high-resolution scans of adults, or with additional low-resolution scans of children, can overcome the resolution gap without introducing a systematic segmentation bias towards adult-like brain volumes in high-resolution images. 
 
\begin{figure}[b]
    \centering
    \includegraphics[width=0.9\columnwidth]{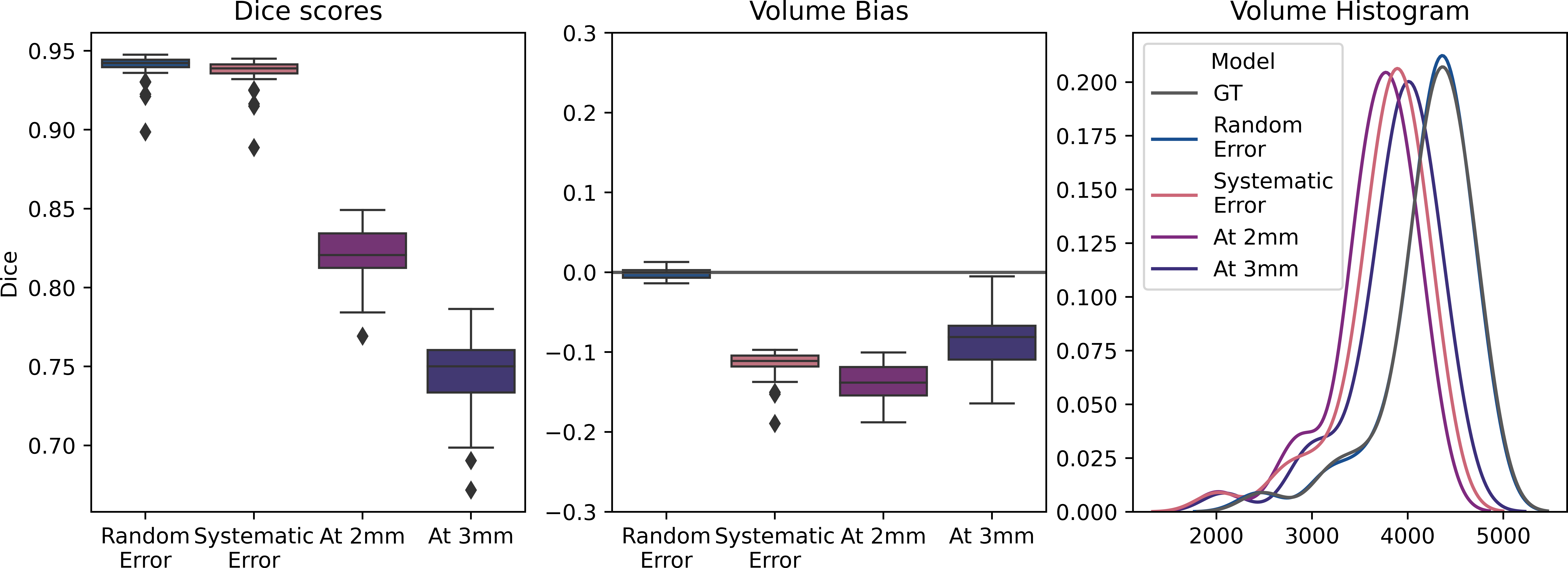}
    \caption{Simple simulations of Random and Systematic (oversegmentation) Errors illustrate significant distribution shifts despite similar Dice scores. In contrast, Volume Bias highlights this error type. Downsampling (to 2/3 mm) results in undersegmentation. }
    \label{fig:intro}
\end{figure}

The analysis of bias in deep learning (DL) for medical imaging is still in its infancy despite its relevance for fairness \cite{Chen2021}. With no established evaluation metrics, recent work employs overlap measures to evaluate racial (cardiac segmentation) \cite{BiasMiccai21,PuyolEuHeart} and motion bias (brain lesion segmentation) \cite{LesionBias}. However, overlap measures (e.g.\ Dice) only detect reduced performance without differentiation between random and systematic errors, as simple simulations reveal. \Cref{fig:intro} illustrates how similar Dice performance can lead to significantly different volume estimation (shift of the volume distribution in the histogram). Systematic volumetry changes often relate to group differentiations and are a far more robust marker for systematic errors than overlap measures. Therefore, the analysis of bias in traditional neuroimaging pipelines have included signed volume measures to capture consistent over- or undersegmentation biases \new{\cite{akudjedu2018,herten_accuracy_2019,de_bruijne_relationship_2021}}. No work has addressed segmentation bias in DL for neuroanatomical segmentations or multiple resolutions. In a multi-resolution setting, care has to be taken to further differentiate methodological from resolution biases. The latter arise when the spatial representation of a structure can not be captured due to the technical resolution limits, i.e.\ the voxel size exceeds the size of the structure itself. Tissue borders in particular are prone to get lost at lower resolutions. Our analysis is therefore based on both, accuracy and volume bias to effectively detect systematic segmentation errors.

The multi-resolution question allows two general training strategies - using either a dedicated fixed-resolution network working on (resampled) images or a network that accepts multiple resolutions during training. We evaluate both strategies and show that: i) single resolution networks fail to adequately capture group differences across resolutions and work only on the original distribution they were trained on, ii) training with resampled images alone is not enough to remove this bias, and iii) multi-resolution approaches, i.e.\ scaling augmentation or resolution independent networks, effectively translate high-resolutional information across the resolution differences and combat the detected bias in segmentation networks.

%evaluate systematic segmentation biases across resolutions in two segmentation tasks and
\section{Methodology}
\begin{figure}
    \centering
    \includegraphics[page=2,clip, trim=0 5.95cm 0 0, width=0.9\columnwidth]{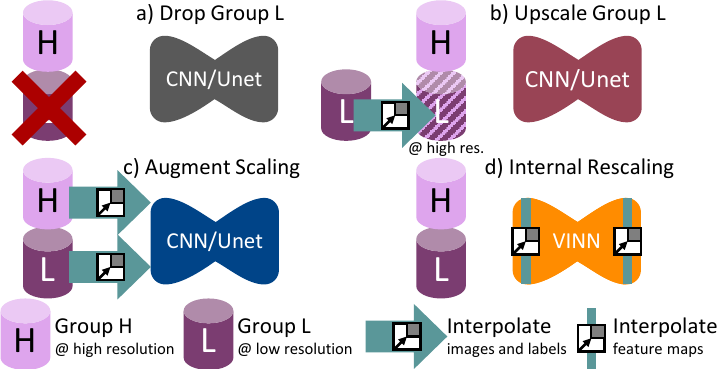}
    \caption{Four approaches to harmonize the resolution-biased datasets for segmentation network: a) drop the lower resolution data, b) upsample the low-resolution dataset to the higher resolution, c) randomly rescale both datasets as augmentation, and d) incorporate a resampling to the higher resolution into the network.}
    \label{fig:methods}
\end{figure}

We explore four approaches to address 
unbiased cross-resolution generalization (see \Cref{fig:methods}): %. Across segmentation tasks, we consider two participant groups for which training data is only available at differing image resolutions. For group H, images we have images and labels at the preferred high resolution and for group L at the deprecated lower resolution. We are specifically interested how this resolution-bias in the data distribution we propagates to systematically biased predictions for group L at higher resolutions. Hence, four approaches for training arise (see \Cref{fig:methods}): 
For approach a), group L (at lower resolutions) is omitted from network training. We expect this reduced training dataset diversity will result in low generalization performance for the left-out group L, if the underlying distributions differ significantly (out of distribution effect). Two approaches combine both groups (H and L) for training by image resampling \new{via bilinear interpolation} (and lossy label interpolation \new{via majority voting}) outside the network: b) upsampling of group L to the higher resolution, or c) scale augmentation during network training. Finally, d) Voxel-size Independent Neural Networks (VINN) \cite{FastSurferVINN} achieve resolution independence by internal rescaling shifting the interpolation step into the network itself. This avoids lossy label interpolation while preserving resolution-independence in a multi-resolution training set-up. While approaches (a) and (b) only operate on a single-resolution, (c) and (d) employ a multi-resolution setting.

\subsection{Networks}
Approaches a) to c) use a classical convolutional neural network (CNN) with scale transitions via maxpooling and index unpooling operations. The VINN architecture \cite{FastSurferVINN} (approach d), on the other hand, implements a flexible network-integrated interpolation for the first scale transition. After this interpolation, the feature voxels are unified to the ``higher resolution''. To isolate architectural choices from these four approaches, we keep the individual layers, blocks, the number of layers and the number of parameters fixed across the architectures in settings a) to d). \new{The code for all models is publicly available on github (\url{github.com/deep-mi/FastSurfer})}. %As in FastSurferVINN, 
The common UNet-like architecture consists of competitive dense blocks with four 3$\times$3 convolutions, batch-normalization and probabilistic rectified linear unit \cite{FastSurferVINN}. Instead of concatenations, the connecting local skip connections employ feature competition through maxout \cite{maxout}. For each approach, we train one network on 2D slices per anatomical plane. During inference, we aggregate per-plane probabilities to better exploit the 3D context. \new{This 2.5D approach has recently been shown to outperform state-of-the-art 3D approaches for whole brain segmentation \cite{roy2022}}. 

\section{Experiments}\label{sec:experiments}
To ensure the presence of differences between group H and L, we select two neuroimaging segmentation tasks with known group differences: Cortical segmentation in adults and children (Task 1) and hippocampal segmentation with groups stratified based on their hippocampal volume (Task 2).

\subsection{Task 1 -- Cortical Segmentation in Adults and Children}
% To introduction
%The cortex is a highly folded structure that is of tremendous interest in neuroimaging. As an effect in aging, cortical gray matter thickness slowly reduces and, therefore, differs significantly between adults and children. Hence, the question arises if networks trained solely on available high-resolution scans of adults, or even with additional low-resolution MRIs of children, can overcome the resolution gap without introducing a systematic segmentation bias in future high-resolution MRIs of children. 
To answer whether the inclusion of low-resolutional children can overcome the group bias in high resolutional evaluation, we construct a dataset with \new{90} participants (\new{45} adults, \new{45} children). The manual cortical segmentations \new{and corresponding T1-weighted intensity images} for the adults group ($\geq$ 20 years) originate from the open-source Mindboggle-101 \cite{Klein2012} dataset. From the UC-Davis MIND Institute Autism Phenome Project \cite{nordahl_autism_2022}, we obtain manually corrected cortical segmentations \new{and corresponding T1-weighted intensity images} from non-autistic children (\new{2.3 to} 4.5 years). The total set of \new{90} scans is split into 50/10/30 for training/validation/testing, each balanced between age groups. 

\subsection{Task 2 -- Stratified Hippocampal Segmentation}
%A strong group separation in the manual labels is a prerequisite to examine bias errors in segmentation networks. 
Using manual hippocampus segmentations \new{and the corresponding T1-weighted intensity images} from the Harmonized Protocol (HarP) \cite{Mueller2005,Boccardi2015}, the MICCAI 2012 Multi-atlas labeling challenge (MALC) \cite{landman2012miccai} and the Internet Brain Segmentation Repository (IBSR), provided by the Center for Morphometric Analysis (CMA) at Massachusetts General Hospital \cite{IBSR}, we construct two groups by selecting the 40 smallest (all either AD or MCI) and 40 largest (dominated by controls) hippocampi. Keeping volume groups balanced, we split them into training/validation/testing sets (n=50/10/20). Akin to Experiment 1, small hippocampi images (group L) are downsampled to a low resolution before training, while images from group H (large hippocampi) reside at a higher resolution. 

%During inference, the capabilities of the networks to segment high-resolution scans is evaluated with respect to systematic biases between the two volume groups.

To analyze the performance of approaches a) to d) in multiple scenarios including different structure sizes, we train networks for three high/low resolution pairs: 1.0/1.4mm, 2.0/3.0mm, and 3.0/4.0 mm. %resample the training and validation set of group L from the native to a lower resolution. 
During inference, however, we always evaluate on the high resolution scans for both groups (H and L) to establish whether the networks generalize to the higher resolution for group L despite training on low-resolution images. 

% Currently, all publicly available manual labels for brain segmentation reside at approximately 1 mm isotropic resolution. Since downsampling of the images only reduces the provided information content per voxel, we resize the images and compare biases in three lower resolution settings: 1.0 mm versus 1.4 mm, 2.0 mm versus 3.0 mm, and 3.0 mm versus 4.0 mm. This represents a scaling range of 1.3 to 1.5 which is comparable to current acquisitions of high-resolution images in relation to standard 1.0 mm. All final metric comparisons are performed at the original high-resolution of the manual labels (i.e.\ 1.0 mm). 

\subsection{Evaluation Measures}
Based on manual segmentations and volumes at 1 mm, we i) compare the Dice Similarity Coefficient (DSC) , ii) the systematic segmentation bias, and iii) the effect of bias on group classification. 
Since DSC cannot differentiate between random and systematic errors, it only serves as a measure of general performance. To quantify the bias, signed differences of a target statistic are required. 
\subsubsection{Systematic Segmentation Bias}
Choosing the volume as the target statistic, we define the (normalized) Volume Bias as the (normalized) difference between predicted (Volume$(f(x))$) and ground truth volume (Volume$(y)$).
\begin{equation}
\text{Volume Bias}(\text{Group}~ i) = E_{(x, y)\, \sim\, \text{Group}~ i} \left [\frac{\text{Volume}(f(x)) - \text{Volume}(y)}{\text{Volume}(y)}\right ]
\label{eq:1}
\end{equation}
A value of zero indicates an unbiased estimate with $>0$ implying over- and $<0$ under-segmentation. To eliminate the contribution of random errors, we form an expected value over the analyzed group. Since our group-wise test sets contain only 15 \new{(Task 1) or 10 (Task 2)} participants, we estimate the expected value by the median and additionally plot its distribution.

\subsubsection{Group Classification}
We use the area under the Receiver operating characteristics curve (AUC) to investigate the effect of  segmentation bias on down-stream analysis. \new{Overall, the task explores, whether ``information'' of known volumetry effects is obfuscated by (training/dataset) bias which is highly relevant for group analysis}. Here, we calculate the AUC for every possible discrimination threshold of the gray matter volumetry with the correct age group assignment (children versus adults) as the predictor variable of interest.

%\section{Experimental Results}\label{sec:results}

\subsection{Training}
We train all networks under equal hardware, hyper-parameter and dataset settings unless stated otherwise. \new{Hyper-parameters were not changed from the original publications except for the internal resolution of VINN \cite{FastSurferVINN} (set to the higher image resolution for all tasks)}. Independent models are trained for 70 epochs with mini batch size of 16 using the AdamW optimizer \cite{adamW}, a cosine annealing learning rate schedule \cite{cosine}, an initial learning rate of 0.001 and momentum of 0.95. The number of epochs between two warm restarts is initially set to 10 and subsequently increased by a factor of two. \new{Based on the validation set, we select the best training state per network and assured convergence of all models.}

\section{Results}
\begin{figure}
    \centering
    \includegraphics[ width=\columnwidth]{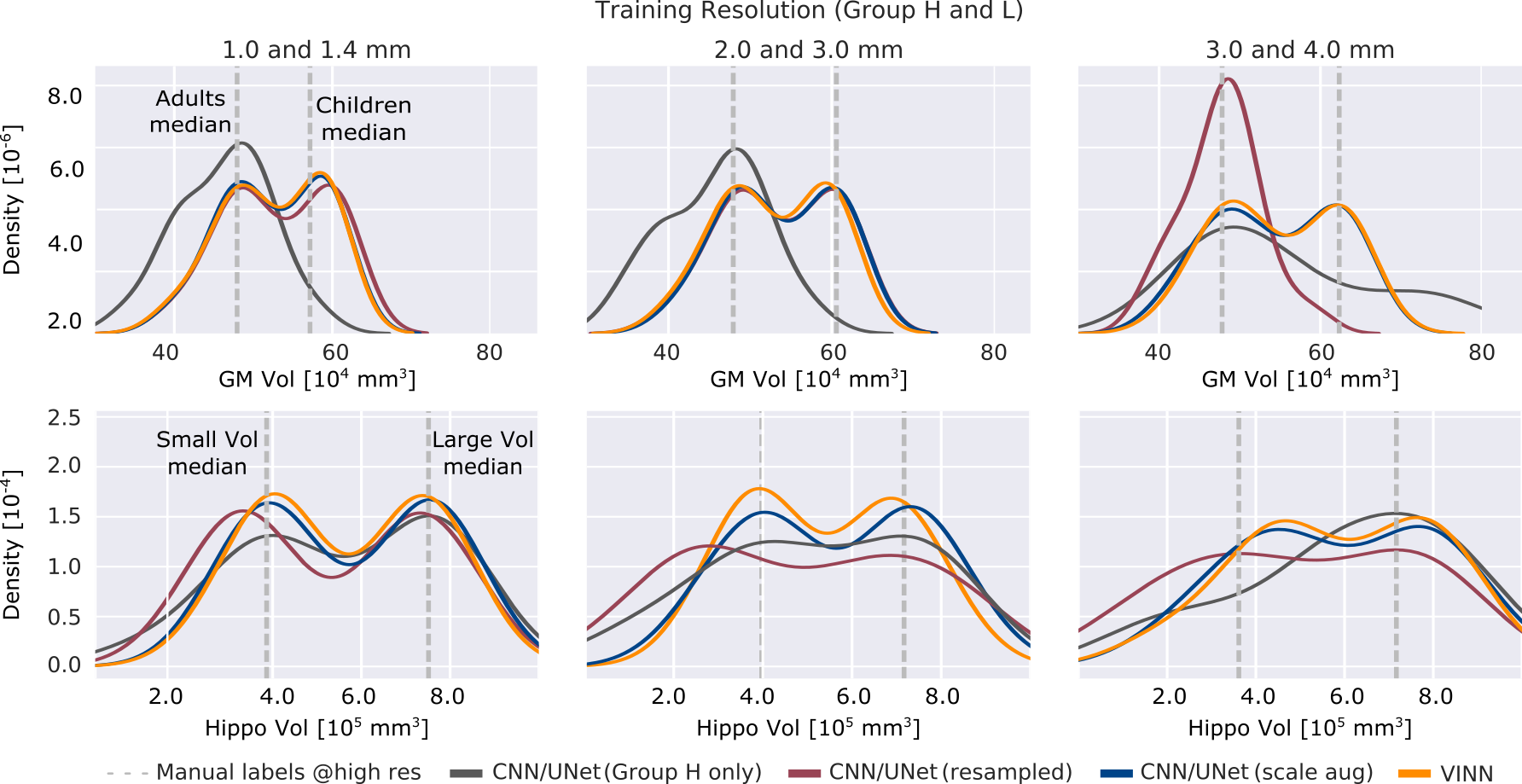}
    \caption{Segmentation bias in single resolution networks. CNN/UNet (black) trained on group H (Adults) only fails to replicate modes of volume distributions (vertical lines) in out of distribution samples (Children). Multi-resolution networks (CNN + scale augmentation, blue and VINN, yellow) efficiently transfer information during training across resolutions and match underlying distribution of ground truth labels.}
    \label{fig:hist}
\end{figure}

First, we analyze how well the different approaches can reproduce the modes of volume distributions for both tasks. \Cref{fig:hist} combines the distributions from multiple experiments to show three origins for systematic differences: 1.\ the different groups, 2.\ the approach used for prediction (method bias), and 3.\ the resolution of the analysis (resolution bias). \Cref{fig:hist} clearly illustrates that approaches (b)-(d) are able to separate the different modes of the two groups in most experiments. The prominent exception is ``CNN/UNet (Group H only)'' (black line). For Task 1 (top row), the Children distribution is basically non existent thus confirming that segmentation networks do not automatically generalize to unseen distributions. Similarly, for Task 2 (bottom row), the distribution for small volumes shifts to the right with decreasing resolutions. A similar trend can be observed for the resampled images ``CNN/UNet (resampled)'' (red line) - the volume distribution shifts to the left and converges to a uni-modal distribution.

% Table takes too much space?
\begin{figure}
    \centering
    \includegraphics[ width=\columnwidth]{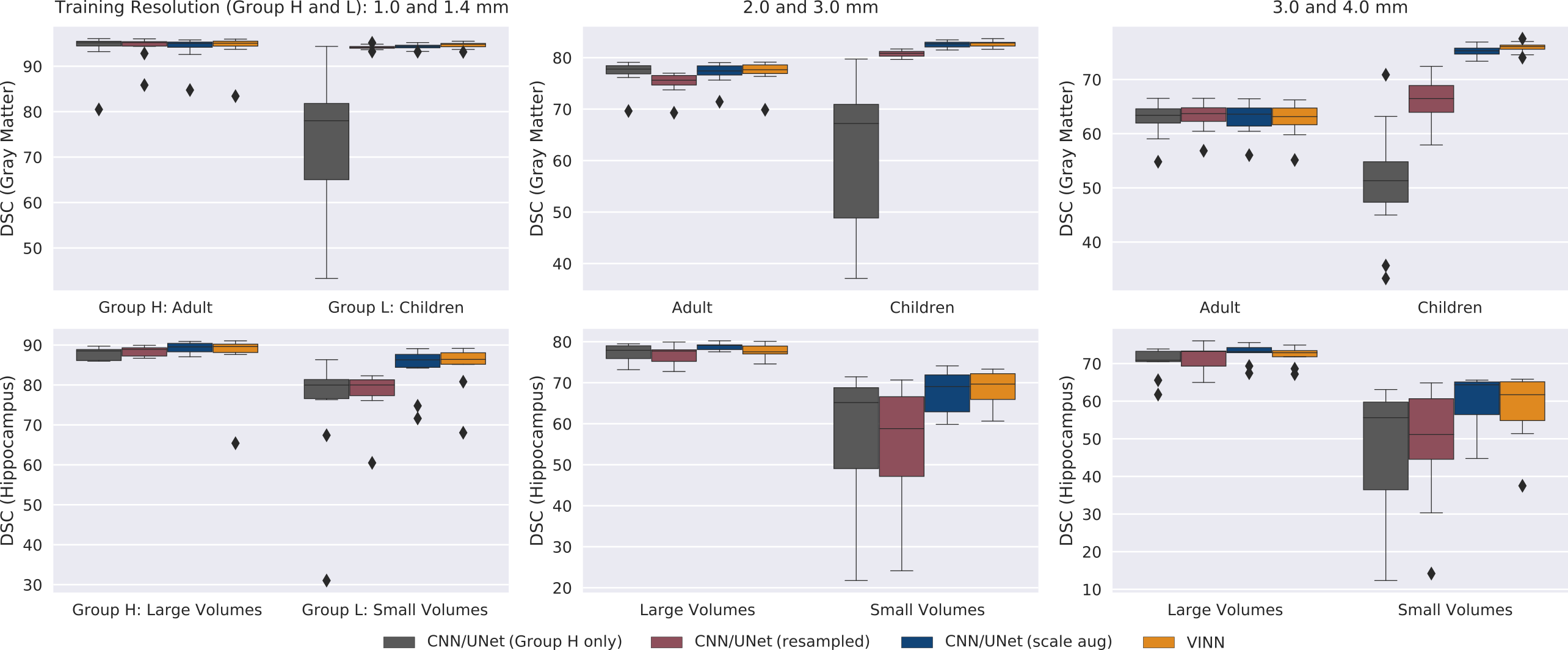}
    \caption{Dice Similarity Coefficient of high-resolution predictions. All networks reach a high DSC for group H (Top: Adults, Bottom: Large Volumes). Accuracy of cortical segmentations with scale augmentation (blue box) or VINN (yellow box) stays high for group L as well (Top: Children, Bottom: Small Volumes), while networks trained on group H only (black box) or on resampled images of both groups (red box) results in reduced DSC values.}
    \label{fig:dsc}
\end{figure}

The findings from the histogram distribution are further substantiated by the calculated accuracy measures (\Cref{fig:dsc}). First, all approaches perform equally well on the high-resolution group encountered during training (group H, Adult and Large Volumes). The drop in DSC to 63 for the low resolution pair (3.0 mm versus 4.0 mm) on the gray matter results from lost cortical details in comparison to the 1.0 mm ground truth (i.e.\ deep sulci are not distinguishable by a complete voxel anymore). This problem is less pronounced on the hippocampus due to its compact shape. Second, cross resolution transfer works best for approaches (c) and (d) (i.e.\ CNN + scale augmentation (blue box) or VINN (yellow box)). The DSC reaches the highest value for both experiments on the left-out high-resolution group L (children and small volumes) and furthermore stays relatively constant across the different resolution pairs with the highest cortical DSC of 94.84, 82.84 and 76.07 for VINN, respectively (hippocampus: 86.4, 69.67, 61.72). As expected, accuracy on group L is lowest for approach (a) (black box). In accordance with results from \Cref{fig:hist}, approach (b) (red box) decreases segmentation performance for small hippocampi.

\begin{figure}
    \centering
    \includegraphics[ width=\columnwidth]{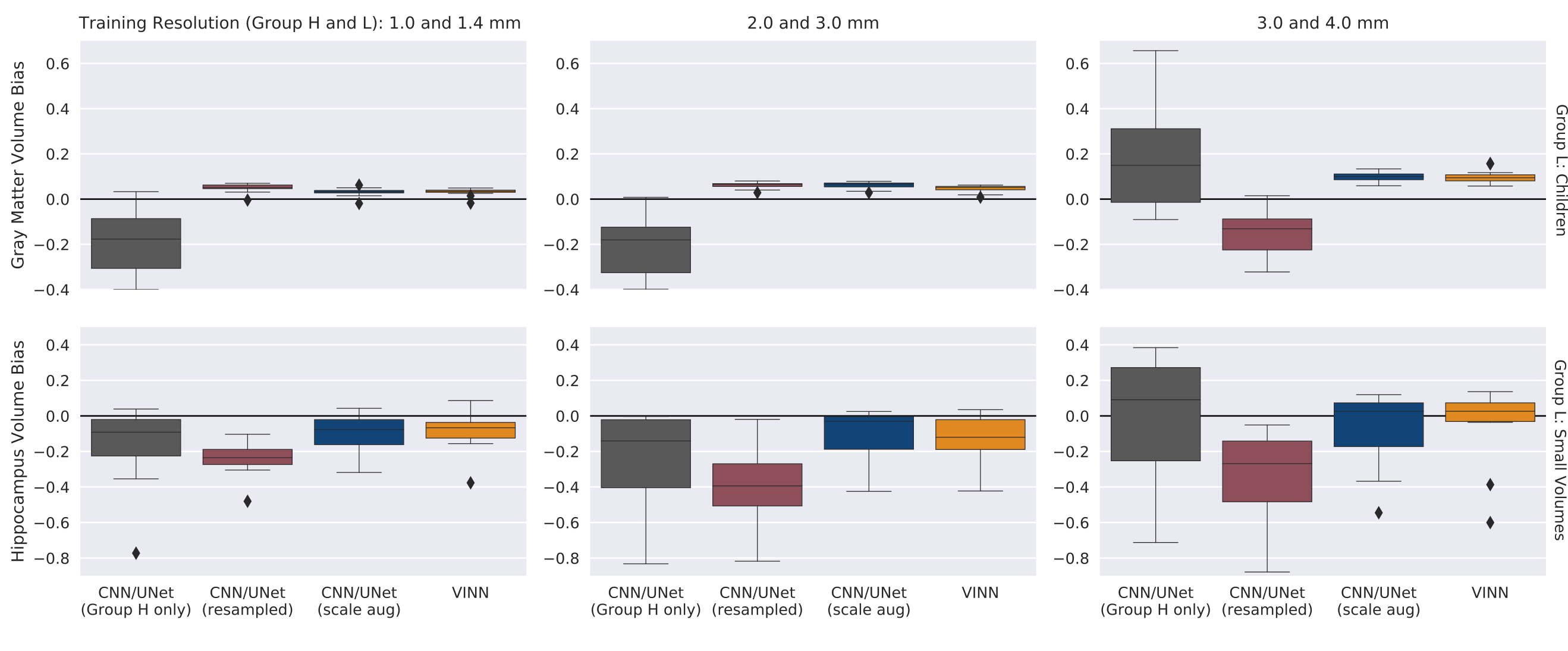}
    \caption{Volume bias in network predictions. Single resolution approaches show systematic biases in the volume estimations of group L (black and red box). Volume estimates based on VINN (yellow box) and CNN with scale augmentation (blue box) change only marginal indicating increased robustness to resolution variances of group L. }
    \label{fig:stats}
\end{figure}
In order to ensure that the observed accuracy drop does indeed point towards a systematic bias in the network predictions, we calculate volume error of the cortical and hippocampal structures. For the multi-resolution approaches (c) and (d) (scale augmentation and VINN), the cortex segmentation of children at high-resolution MRI is consistent with a median volume error close to 0 (\Cref{fig:stats}, blue and yellow box). The single resolution approache (a) consistently underestimates the cortical gray matter volume for resolutions 1.0 and 2.0 mm by around 20 \% (black box). Approach (b) shows a bias with shifts to lower resolutions (red box, 13 \% at 3.0 versus 4.0 mm). This effect is increasingly apparent on the high-resolution participants with small hippocampi in experiment 2. Here, the resampling leads to a constant underestimation of the hippocampal volume by 23.5 to 39 \%. Multi-resolution approaches (c) and (d) are again more consistent with a variance in volume measures across resolutions between only 2 and 6 \%.

Taken together, the results point towards a systematic accuracy bias in single resolution segmentation networks. Next, we evaluate if the detected bias is strong enough to carry over to down-stream analysis. The incompetence of the single resolution approach (a) to translate group differences across resolutions clearly reduces classification performance (\Cref{tab:auc}, AUC of 0.36). The method bias hence actively undermines group separation. At 3.0 mm, the AUC recovers probably because the failure to estimate volumes at all rather than the actual volume value becomes the basis for the decision threshold. The accuracy results in \Cref{fig:dsc} corroborate this hypothesis. Interestingly, the volume bias also translates to a classification bias for approach (b) at the lower resolutions (3.0 mm). Here, the AUC drops by more than 50 \% to 0.4 when attempting to classify children versus adults. Only scale augmentation and VINN preserve a high AUC of above 0.95 across all resolution pairs. 

\begin{table}[hb!]
\begin{tabular}{c|
>{\columncolor[HTML]{EFEFEF}}c c
>{\columncolor[HTML]{EFEFEF}}c c
>{\columncolor[HTML]{EFEFEF}}c }
\multicolumn{1}{c|}{\textbf{\begin{tabular}[c]{@{}c@{}}Training Res \\ Group H \& L\end{tabular}}} & \multicolumn{1}{c}{\cellcolor[HTML]{EFEFEF}\textbf{\begin{tabular}[c]{@{}c@{}}CNN/UNet \\ (Group H only)\end{tabular}}} & \multicolumn{1}{c}{\textbf{\begin{tabular}[c]{@{}c@{}}CNN/UNet \\ (resampled)\end{tabular}}} & \multicolumn{1}{c}{\cellcolor[HTML]{EFEFEF}\textbf{\begin{tabular}[c]{@{}c@{}}CNN/UNet \\ (scale aug)\end{tabular}}} & \textbf{VINN} & \multicolumn{1}{c}{\cellcolor[HTML]{EFEFEF}\textbf{\begin{tabular}[c]{@{}c@{}}Manual \\ labels\end{tabular}}} \\ \hline
\textbf{1.0 \& 1.4 mm}                                                                             & \textbf{0.36}                                                                                                           & 0.93                                                                                         & 0.92                                                                                                                 & 0.92          & 0.90                                                                                                          \\
\textbf{2.0 \& 3.0 mm}                                                                             & \textbf{0.35}                                                                                                           & 0.95                                                                                         & 0.94                                                                                                                 & 0.94          & 0.90                                                                                                          \\
\textbf{3.0 \& 4.0 mm}                                                                             & 0.89                                                                                                                    & \textbf{0.4}                                                                                 & 0.96                                                                                                                 & 0.96          & 0.90                                                                                                         
\end{tabular}
\caption{Area under curve (AUC) for classification of "Children" based on volumetry information. Biased segmentations from CNN/UNet significantly decrease classification performance. Scale augmentation and VINN successfully diminish biases across resolutions and achieve AUCs above 0.92. }
    \label{tab:auc}
\end{table}


\section{Conclusion}
Overall, we provide the first comprehensive analysis of bias between groups in segmentation networks. We show that single resolution networks work well on their training distribution but fail to generalize group differences across resolutions. The detected volume bias further propagates to classification tasks. Finally, we show that scale augmentation in CNNs as well as alternative architectures with build-in resolution independence like VINN can help reduce these biases and effectively translate information from high- to lower-resolution scans. The findings are highly relevant to all fields dealing with resolution differences. Care should specifically be taken, when considering future high-resolution image acquisitions: subsequent processing  with single-resolution deep-learning based analysis pipelines should be avoided. 

\subsubsection{Acknowledgements:} This work was supported by the Federal Ministry of Education and Research of Germany (031L0206), by NIH (R01 LM012719, R01 AG064027, R56 MH121426, and P41 EB030006), and the Helmholtz Foundation (project DeGen).
Funding for the APP dataset was provided by NIH and UC Davis MIND Institute. The study team would like to acknowledge Devani Cordero, of the Martinos Center of Biomedical Engineering and thank all of the families and children for their participation. 
%Funding for the Autism Phenome Project was provided by the National Institute of Mental Health (R01MH104438, R01MH103284, R01MH103371), the MIND Institute Intellectual and Developmental Disabilities Research Center (P50 HD103526) and the Autism Center of Excellence (P50HD093079) awarded by the National Institute of Child Health and Development. The study team would like to acknowledge Devani Cordero, of the Martinos Center of Biomedical Engineering and thank all of the families and children for their participation. 
Further, we thank and acknowledge the providers of the datasets listed in \Cref{sec:experiments}.

\bibliographystyle{splncs04unsrt}
\bibliography{paper1422.bib}
%Author, F.: Article title. Journal \textbf{2}(5), 99--110 %(2016)
%Author, F., Author, S.: Title of a proceedings paper. In: %Editor,
%F., Editor, S. (eds.) CONFERENCE 2016, LNCS, vol. 9999, pp. %1--13.
%Springer, Heidelberg (2016). \doi{10.10007/1234567890}
%Author, F., Author, S., Author, T.: Book title. 2nd edn. %Publisher,
%Author, A.-B.: Contribution title. In: 9th International %Proceedings
%on Proceedings, pp. 1--2. Publisher, Location (2010)
%LNCS Homepage, \url{http://www.springer.com/lncs}. Last %accessed 4
%\end{thebibliography}
\end{document}